\documentclass[%
 aip,
 amsmath,amssymb,
 reprint,%
]{revtex4-1}

\usepackage{graphicx}% Include figure files
\usepackage{dcolumn}% Align table columns on decimal point
\usepackage{bm}% bold math
\usepackage[utf8]{inputenc}
\usepackage[T1]{fontenc}
\usepackage{mathptmx}
\usepackage{etoolbox}

\usepackage[justification=raggedright]{caption}
\usepackage{subcaption}
\usepackage{multirow}
\usepackage{makecell}

\makeatletter
\def\@email#1#2{%
 \endgroup
 \patchcmd{\titleblock@produce}
  {\frontmatter@RRAPformat}
  {\frontmatter@RRAPformat{\produce@RRAP{*#1\href{mailto:#2}{#2}}}\frontmatter@RRAPformat}
  {}{}
}%
\makeatother
\begin{document}

\preprint{AIP/123-QED}

\title[Experimental investigation of plasma-electrode interactions on the ZaP-HD sheared-flow-stabilized Z-pinch device]{Experimental investigation of plasma-electrode interactions on the ZaP-HD sheared-flow-stabilized Z-pinch device}
% Force line breaks with \\
\author{A. A. Khairi}
\author{U. Shumlak}%
 \email{aqilk@uw.edu.}
\affiliation{Aerospace and Energetics Research Program, University of Washington, Seattle, WA 98195-2400%\\This line break forced with \textbackslash\textbackslash
}%

\date{\today}% It is always \today, today,

\begin{abstract}
The ZaP-HD sheared-flow-stabilized (SFS) Z-pinch device is a testbed for experimental investigation of plasma-electrode interactions. The graphite electrode is exposed to a high temperature, high density Z-pinch plasma while supplying large pinch currents. In-situ measurements of the gross carbon erosion flux obtained with S/XB spectroscopy exceed the expected flux from physical sputtering, but have reasonable agreement with the expected sublimation flux. Comparison of the ionization mean free paths of neutrals produced through both erosion processes shows that sublimated carbon is ionized within the sheath while sputtered carbon is ionized beyond the sheath. This suggests a process of electrode recycling and self-healing through redeposition. The sputtered carbon is primarily responsible for net erosion. Ex-situ analysis of electrode material is enabled by the design of a removable coupon. Three different plasma exposure conditions varied the pinch current and number of pulses. Net mass loss measurements support the physical picture of electrode recycling. Erosion rates range from 0.01 to 0.1~mg/C, which are comparable to existing arc discharge devices. Measurements of the microscopic surface morphology and roughness reveal irregular consolidated structures and general smoothing except at high particle fluence. Crack formation suggests the importance of repetitive thermal cycles. Definitive features of sputtering such as pitting and cratering are absent, although further study is needed to attribute the observed changes to other processes. These results indicate some alignment with erosion processes in high-powered arc discharges, which successfully operate solid electrodes in extreme environments. This provides confidence in managing electrode erosion in the SFS Z-pinch configuration.
\end{abstract}

\maketitle

\section{\label{sec:intro}Introduction}
% Paragraph 1 - Broad relevance
The interaction of plasmas and materials is a fundamental process that occurs throughout plasma physics. In space, these plasma-material interactions (PMI) are relevant to the study of star formation in the interstellar medium \cite{Bringa_2007} and interactions of the ionosphere with space debris \cite{SEN_2015} and spacecraft materials. \cite{Engelhart_2018} In artificial plasmas, these interactions have been harnessed for electrical lighting, \cite{Brush_1881, Waymouth_1991} biomedicine, \cite{Suschek_2024} and food safety. \cite{Pankaj_2018} Perhaps most prominent are the impressive array of plasma processing techniques \cite{plasmaprocessing1991} used in semiconductor manufacturing and other industries foundational to the modern material and digital world.

% Motivation for studying plasma-electrode interactions
In magnetic fusion, PMI is an inevitable feature of all confinement schemes. The formidable heat and particle fluxes generated by a fusion-grade plasma present a critical challenge in the development of materials for fusion devices. The resulting interactions are varied and complex, affecting both the material condition and the plasma performance. Although the importance of PMI in fusion is well-known and the subject of decades of research, \cite{Zinkle2013, Linsmeier_2017, Linke_2019_MRE} much of this work has focused on toroidal configurations such as the tokamak and stellarator. In these devices, plasma-facing components (PFCs) such as the first wall and the divertor are situated at the edge of the confined plasma and do not drive any plasma current. While these components must still withstand considerable particle and heat fluxes, they are distinct from electrodes which contact plasma and supply the current. A solid electrode in direct contact with the core plasma is a defining feature of the sheared-flow-stabilized (SFS) Z pinch configuration. The resulting plasma-electrode interactions are not fully understood, especially at the high current densities and heat fluxes necessary for a Z-pinch fusion power plant, where their importance is expected to increase. \cite{Shumlak2020} Magnetic effects such as prompt redeposition \cite{Naujoks_1997} and plasma fluxes across field lines tangential to a surface \cite{Stangeby_1992} are important factors to consider, although are not the focus of this paper. Operation of a solid electrode in extreme environments also occurs in vacuum and atmospheric arc discharges \cite{Kimblin_1974}, cathodic arcs, \cite{NG_2014} arc furnaces, \cite{Sævarsdóttir_2016} and plasma-based propulsion. \cite{Brown_2020, OHAIR_1989} These systems span a wide range of current densities but are typically characterized by low plasma temperatures and therefore lower heat fluxes. Ultimately, electrode erosion limits component lifetime and operational capacity of the SFS Z pinch, both critical factors in the development of a fusion power plant. \cite{Thompson_2023_FST} This motivates an investigation into the physical processes that govern electrode erosion, which will inform the design of more robust electrodes and optimize operational parameters to improve electrode longevity.

% Discuss sublimation and sputtering erosion mechanisms, brief lit review
Erosion can arise from various physical mechanisms associated with particle and thermal loading of the PFC. Particle flux from the plasma leads to bombardment of the solid material by energetic ions which can eject particles from the surface in a phenomenon known as physical sputtering. Physical sputtering is an important erosion mechanism for tokamak PFCs, \cite{BROOKS2011S112, Abrams_2019} motivating dedicated study using high-energy ion beams. \cite{SEYEDHABASHI_2025} The heat flux from the plasma, combined with additional resistive heating in the case of an electrode, can lead to elevated surface temperatures that result in melting or sublimation. Sublimation has been observed in graphite electrodes of high energy spark gaps \cite{Gordon_1982} and has been investigated at fusion-relevant heating conditions by laser beam heating. \cite{WU_2002}

% Literature review and gap identification
A number of existing PMI testbeds seek to replicate the edge plasma conditions expected on ITER, such as 20~MWm$^{-2}$ steady-state heat flux, GWm$^{-2}$ transient heating, and low electron temperatures around 10~eV. \cite{Ikeda_2007} While contemporary large tokamaks enable PMI studies for ITER-like operational scenarios and magnetic field configurations, they lack the sufficient heat, ion, and neutron fluxes to fully replicate the expected conditions. \cite{Linsmeier_2017_Facilities} Linear plasma devices (LPDs) serve to fill this gap with their versatility, steady-state operation capability, and moderate costs. \cite{Kreter2011} Furthermore, these dedicated testbeds provide well-defined exposure conditions and good access to material samples. \cite{Linsmeier_2017_Facilities} In terms of the plasma exposure, the SFS Z pinch produces plasmas with 1 keV electron temperatures \cite{Shumlak2017} and 10 GWm$^{-2}$ heat fluxes. \cite{Thompson_2023_POP} These are incident on a solid electrode that is supplying hundreds of kiloamperes of current. Ion impact energies enhanced by the electrode bias voltage, and a relatively small area of plasma contact cause intense plasma-electrode interactions. However, pulse durations are on the order of microseconds compared to the long duration or steady-state pulses on existing testbeds, necessitating an increased number of pulses for comparable particle fluences.

% High-level approach of described research, novel contributions, findings, and proposed explanation.
This paper presents the results of an experimental investigation of the plasma-electrode interactions on the ZaP-HD SFS Z-pinch device. \cite{Shumlak2017} The graphite electrode is subjected to a high temperature, high density Z-pinch plasma, while supplying pinch currents up to 136 kA. Gross electrode erosion is measured in-situ and surface morphology changes are measured with ex-situ diagnostics. These represent the first measurements of the electrode material response to plasma exposure in the SFS Z-pinch configuration, which indicate the relative importance of multiple erosion mechanisms, and suggest a mode of operation that promotes electrode recycling and self-healing.

% Paper organization
The paper is organized as follows. In Sec.~\ref{sec:apparatus}, the ZaP-HD experiment, removable electrode coupons, and relevant diagnostic systems are described. Section~\ref{sec:exp1} shows that total erosion measurements are comparable to the expected sublimation flux, and Sec.~\ref{sec:exp2} describes a physical mechanism for recycling carbon that significantly reduces net erosion. Section~\ref{sec:exsitu} presents mass loss measurements, net erosion rates, and the ex-situ surface analysis. A summary of the findings and their implications is given in Sec.~\ref{sec:conc}.

%  ZaP-HD cross-section figure
\begin{figure*}
\includegraphics[scale=0.8]{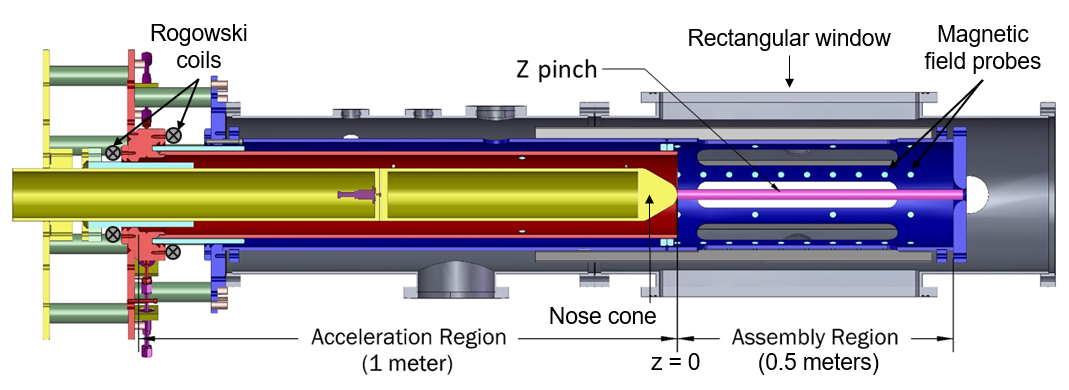}
\caption{\label{fig:zaphd}Cross-sectional machine drawing of the ZaP-HD SFS Z-pinch device illustrating the Acceleration and Assembly Regions. Neutral gas is ionized and accelerated down the Acceleration Region. In the Assembly Region, the plasma forms a Z-pinch configuration and is compressed by the axial current. The Z-pinch plasma (shown schematically in magenta) conducts the discharge current and directly contacts the inner electrode at the nose cone. Four rectangular windows in the Assembly Region provide optical access to the Z-pinch plasma and the nose cone for erosion measurements. Two Rogowski coils measure the total plasma current and the compression current. An array of magnetic field probes in the Assembly Region measures the azimuthal magnetic field produced by the Z pinch.}
\end{figure*}
\section{\label{sec:apparatus}Experimental Apparatus}

This section provides a brief description of Z-pinch formation on the ZaP-HD experiment as relevant to this study. Detailed discussion of SFS Z-pinch formation has been presented in previous publications. \cite{Shumlak_2001_PRL, Golingo_2005, Shumlak_2009, Shumlak2012_FST, Shumlak2017} In addition, this section discusses the implementation of removable electrode coupons, relevant diagnostics, and parameters of the experimental campaign.

\subsection{\label{sec:zaphd}ZaP-HD Experiment}

The ZaP-HD experiment at the University of Washington (UW) investigates sheared-flow stabilization of Z-pinch plasmas at high temperature and density. Earlier theoretical predictions \cite{Shumlak1995} of Z-pinch stability from sheared flows was supported by investigations on the ZaP experiment.
\cite{Shumlak_2001_PRL, Golingo_2005, Shumlak_2009} ZaP-HD explores scaling of plasma parameters predicted by analysis of the radial force balance between the plasma pressure and the magnetic field,
\begin{equation}
    \frac{d}{dr}\left (n_ik_BT_i + n_ek_BT_e \right) = -\frac{B_\theta}{\mu_0r}\frac{d}{dr}\left( rB_\theta \right),
    \label{radeq}
\end{equation}
where $n_i$ and $n_e$ are the ion and electron number densities, $T_i$ and $T_e$ are the ion and electron temperatures, and $B_\theta$ is the self-generated azimuthal magnetic field. Scaling relations derived from this equilibrium indicate higher plasma parameters can be achieved in a compact device by increasing the pinch current, \cite{Shumlak2012_FST, Shumlak2017, Shumlak2020} which has motivated development for commercial fusion applications. \cite{Levitt_2023}

The ZaP-HD experiment provides enhanced control of flow Z-pinch formation by combining a coaxial accelerator with a Z-pinch assembly region. This is illustrated in Fig.~\ref{fig:zaphd}. A capacitor bank is discharged across the inner (yellow) and middle (red) electrodes in the Acceleration Region for ionization and acceleration of neutral gas injected into the annular volume. The Lorentz force accelerates the plasma downstream until it reaches the end of the inner electrode, where it assembles into a Z-pinch configuration, represented schematically in magenta. In the Assembly Region, a second capacitor bank is discharged across the inner and outer (blue) electrodes, driving additional current that compresses the plasma on axis. Sheared flow is maintained by the supply of axially flowing plasma from the Acceleration Region. This decoupling of plasma acceleration and compression processes enables larger pinch currents to be driven. ZaP-HD has demonstrated 1 keV electron temperatures with an electron density of 10$^{24}$ m$^{-3}$, and maintained stability for quiescent periods of 50~$\mu$s. \cite{Shumlak2017} The Z-pinch plasma conducts current from the outer electrode to the inner electrode, making direct contact with the inner electrode at the nose cone. This location of plasma-electrode interaction is the focus of this investigation.

\subsection{Removable electrode coupons}
\begin{figure*}
\includegraphics[scale=0.9]{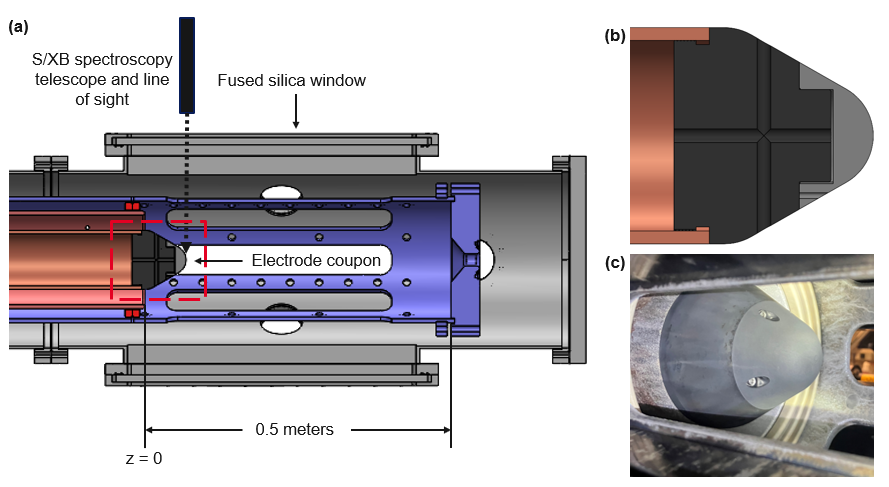}
\caption{\label{fig:pminosecone}(a) Cross-sectional machine drawing of the Assembly Region showing the inner electrode positioned 8~cm downstream of $z=0$, providing optical access for spectroscopy measurements through the fused silica windows. The redesigned electrode assembly incorporating the removable coupon is outlined in red. Removal of the windows enables access for coupon replacement through the slots in the outer electrode. (b) Enlarged section-view of the electrode coupon assembly. Fasteners are not visible in the plane of the cross-section. (c) Image of a pristine electrode coupon installed on ZaP-HD.}
\end{figure*}

Previously, experiments on ZaP-HD studied the plasma-material interaction of graphite targets placed in the downstream plasma. \cite{Forbes_2020_PhD} However, these targets had no appreciable current flow and therefore did not reproduce the plasma-electrode interactions that are the focus of this study. Samples of the electrode material are obtained by redesigning the inner electrode with a removable coupon. Figure~\ref{fig:pminosecone}(a) shows the ZaP-HD Assembly Region with the modified electrode outlined in red. An enlarged view of this region is given in Fig.~\ref{fig:pminosecone}(b). The coupon is effectively the entire tip of the electrode nose cone with internal mating surfaces. Four silver-plated screws, visible in Fig.~\ref{fig:pminosecone}(c), fasten the coupon to the larger graphite base component. All coupons, as well as the base component, were machined out of POCO AXF-5Q graphite. Coupon dimensions, shown in Fig.~\ref{fig:surfmap}, were restricted by the sample stage of the ex-situ surface analysis tools and the slots within the outer electrode shown in Fig.~\ref{fig:pminosecone}(a) which were used for installation. Coupon installation can be completed within two hours, enabling rapid and frequent replacement.
\subsection{Diagnostics}

\begin{figure}
\includegraphics[scale=0.46]{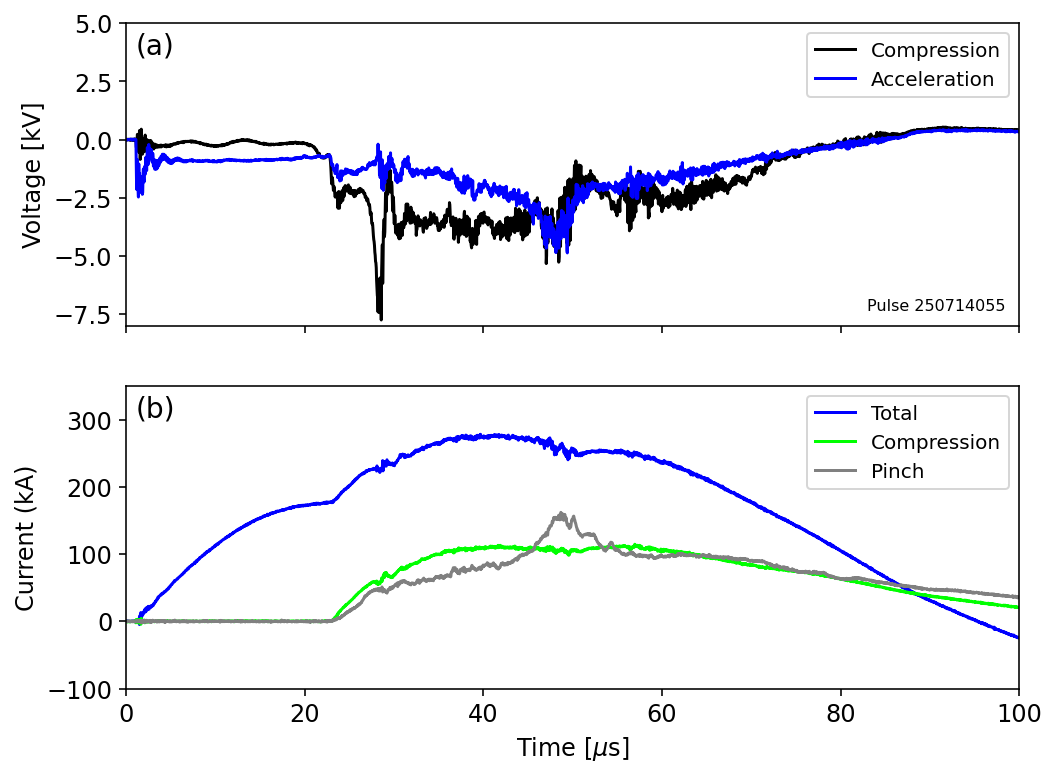}
\caption{\label{fig:typical}Typical traces of (a) voltages and (b) currents recorded during a ZaP-HD pulse. The initial capacitor bank discharge occurs at 2~$\mu$s, applying a voltage to the Acceleration Region and allowing current to flow through the plasma. After a 20 $\mu$s delay, the second capacitor bank discharges, applying a voltage to the Assembly Region that drives an axial current for pinch compression. The pinch current is calculated using Eq.~\ref{eq:pinchcurrent} using the azimuthal array of probes at $z=10$~cm.}
\end{figure}

The ZaP-HD experiment is equipped with various diagnostics to monitor the applied voltages, currents, and the magnetic field. These are described in detail in Ref.~\onlinecite{Forbes_2020_PhD}. A brief explanation is provided here. The applied voltage across each electrode pair is measured by diverting a small amount of the discharge current to high power resistors and measuring the current. A simple calculation with Ohm's law, $V = IR$, provides the voltage measurement. Two Rogowski coils measure the total discharge current and the compression current driven through the Assembly Region. Their positions are labeled in Fig.~\ref{fig:zaphd}. The Rogowski coil enclosing only the inner electrode measures the total current, and the coil enclosing the inner and the middle electrode measures the compression current. Also shown in Fig.~\ref{fig:zaphd} are the magnetic field probes embedded in the outer electrode. These probes are arranged in axial and azimuthal arrays and measure the azimuthal magnetic field in the Assembly Region. The pinch current can be calculated from the magnetic field measurements at a particular axial location by solving Amp\`ere's Law for the enclosed current $I$,
\begin{equation}
    I = \frac{2\pi r_{w}B_{\theta}}{\mu_0},
    \label{eq:pinchcurrent}
\end{equation}
where $B_{\theta}$ is the mean azimuthal magnetic field at some axial position and $r_{w}=0.1$~m is the inner radius of the outer electrode where the probes are mounted. Typical voltage and current traces during a ZaP-HD pulse are shown in Fig.~\ref{fig:typical}. The first capacitor bank discharges across the Acceleration Region at about 2~$\mu$s, causing a rise in the acceleration voltage and total current. After a 20~$\mu$s delay, the second capacitor bank discharges across the Assembly Region, causing the compression voltage and current traces to rise. The pinch current is calculated using Eq.~\ref{eq:pinchcurrent} using the magnetic field probe measurements at $z=10$~cm, which is the position closest to the tip of the inner electrode.
\begin{figure}
\includegraphics[scale=0.52]{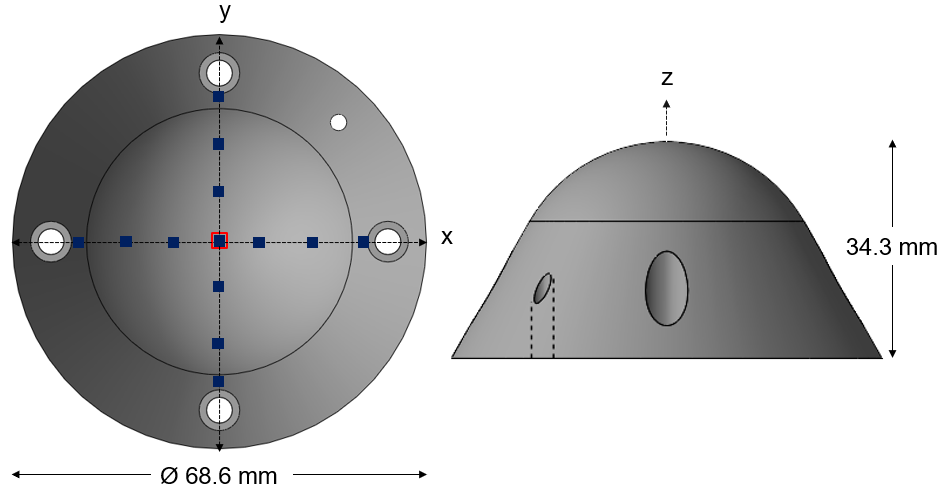}
\caption{\label{fig:surfmap}Top-down and side views of the graphite electrode coupon with major dimensions labeled. SEM measurements (blue squares) were made in 10~mm increments along perpendicular axes, while profilometer measurements (red square) were limited to the center of the coupon.}
\end{figure}

New diagnostics were implemented for the purpose of studying electrode erosion, augmenting the existing diagnostic suite available on ZaP-HD. In-situ measurements of the eroded carbon flux were obtained with S/XB spectroscopy. \cite{Khairi_2025_RSI} In this method, the carbon erosion flux is inferred \cite{Behringer_1989} by measuring the line-integrated photon flux of C-III ion emission at 229.7~nm and applying temperature and density-dependent ionization and excitation rates from the ADAS database. \cite{Summers_2011_ADAS} The position of the electrode was brought forward to $z=8$~cm as shown in Fig.~\ref{fig:pminosecone}(a), from the original position at $z=0$~cm shown in Fig.~\ref{fig:zaphd}. This provides optical access for spectroscopy through the windows.

Ex-situ surface analysis of the electrode coupons was performed with Scanning Electron Microscopy (SEM) and optical profilometry. Figure \ref{fig:surfmap} shows the locations of measurements using the SEM (blue squares) and optical profilometer (red square). Micrographs were obtained along the $x$ and $y$ axes every 10 mm for various magnifications. Profilometry measurements were limited to the tip of the coupon to maintain a useful depth of field.

\subsection{Experimental parameters}
Three experimental campaigns were conducted that varied the conditions of the plasma exposure to the electrode. Two parameters were varied: the applied voltage during compression of the plasma and the number of pulses. The compression voltage controls the pinch current. Higher pinch currents lead to increased density and temperature based on the scaling described in Sec.~\ref{sec:zaphd}. This results in higher current densities and greater particle and heat flux to the electrode. The number of pulses controls the overall duration of plasma exposure. The parameters of each plasma exposure case are summarized in Table~\ref{tab:cases}, along with the maximum pinch current calculated with Eq.~\ref{eq:pinchcurrent} using magnetic field probe measurements at $z=10$~cm. The current values are averaged over all pulses within each campaign. For all cases, the initial capacitor bank discharge for plasma generation and acceleration was set to the maximum charge voltage of 9~kV. Case I represents the most intense plasma exposure in this study, while Case III is the least intense. 
\begin{table}[h!]
\caption{\label{tab:cases}Parameters for the three plasma exposure cases conducted on individual graphite electrode coupons. The compression voltage setting controls the current driven through the pinch during compression in the Assembly Region. The number of pulses controls the total duration of plasma exposure. The maximum pinch current is a campaign-averaged value calculated from magnetic field measurements at $z=10$~cm.}
\begin{ruledtabular}
\begin{tabular}{cccc}
\makecell[c]{Case \text{$\#$}} &
\makecell[c]{Compression voltage \\ \text{[kV]}} &
\makecell[c]{ Total pulses} &
\makecell[c]{Max. pinch current \\ \text{[kA]}}  \\
\hline
% I & Mixed & 155 & $\geq$ 81\\
I & 9 & 200 & 136\\
II & 7 & 50 & 114\\
III & 5 & 42 & 98\\
\end{tabular}
\end{ruledtabular}
\end{table}

\section{\label{sec:experimentalresults}Experimental Results}
\subsection{\label{sec:exp1}Electrode erosion is dominated by sublimation}

\begin{figure}
\includegraphics[scale=0.43]{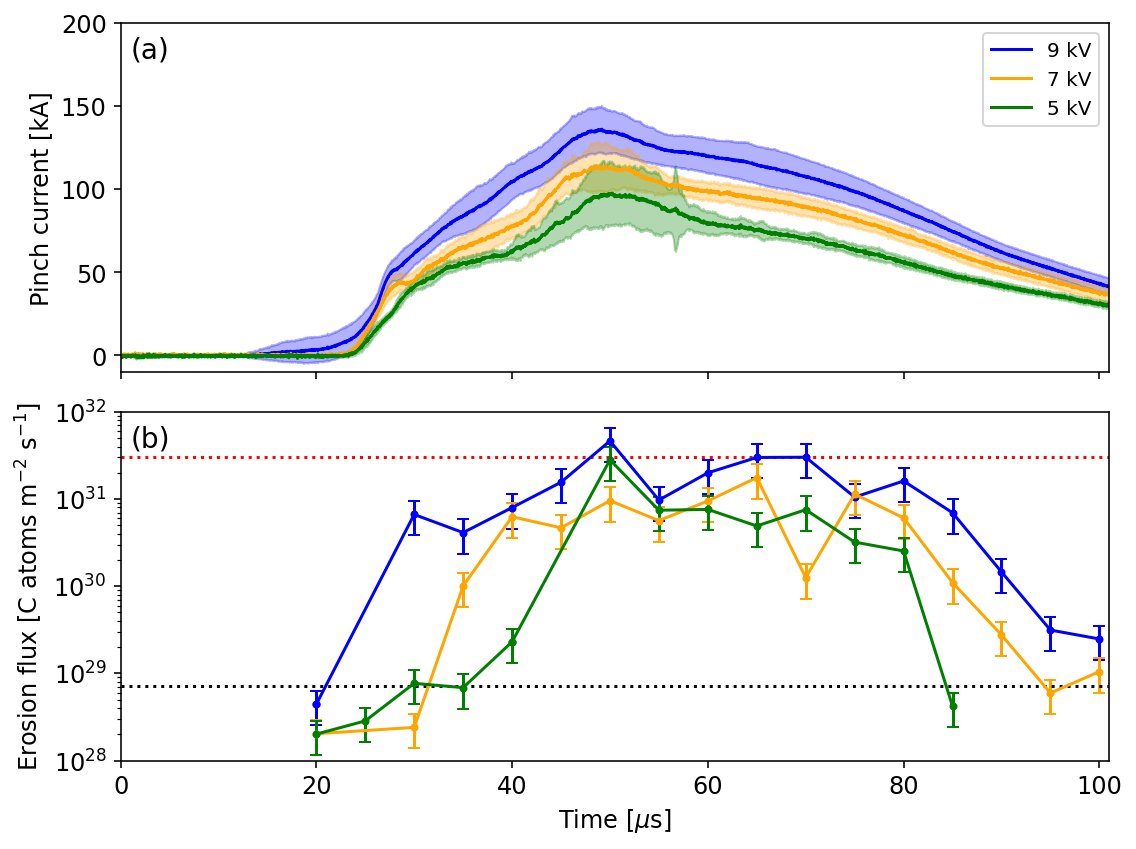}
\caption{\label{fig:sxb_currents} Time evolution of the (a) pinch current and (b) flux of eroded carbon for the compression capacitor bank voltages used in the study. In (a), the shaded regions show the standard deviation of the pinch current over all pulses in each campaign. Increasing this voltage setting increases the pinch current, which corresponds to greater measured erosion flux. The red and black dotted lines correspond to theoretical upper limits of the flux from sublimation and physical sputtering, respectively.}
\label{sxbplot}
\end{figure}

% SXB measurements and comparison to theoretical values
Measurements of the eroded carbon flux from the electrode are observed throughout the duration of the plasma pulse. To obtain a time-resolved measurement, the S/XB diagnostic was triggered at intervals of 5~$\mu$s over repeated pulses. The results are shown in Fig.~\ref{fig:sxb_currents} for the three compression voltage settings. Figure~\ref{fig:sxb_currents}(a) shows the campaign-averaged pinch current measured using the $z=10$~cm probe array. The shaded regions show the standard deviation across all pulses in each campaign. The variation in the pinch current for the 5~kV setting has some overlap with currents measured for the 7~kV setting close to the peak of the trace. Figure~\ref{fig:sxb_currents}(b) shows the corresponding erosion measurements. Each data point is obtained from an individual pulse. The peak pinch current occurs at approximately 50~$\mu$s, which is concurrent with the period of high measured carbon erosion. The particle and heat flux from the plasma should be the largest close to the time of peak pinch current, leading to greater erosion. The erosion measurements show a similar overlap between the 5~kV and 7~kV settings which is attributed to the pinch current overlap.

Figure~\ref{fig:sxb_currents}(b) also compares the measured total erosion with expected values for the sublimation flux (red dotted line) and physical sputtering flux (black dotted line). These values are calculated in Ref.~\onlinecite{Khairi_2025_RSI}, but are summarized here. The sublimation flux $\Gamma_{sub}$ is calculated using the following expression
\begin{equation}
    \Gamma_{sub} = \frac{(2kT_i + \mid eV_{sheath}\mid)\Gamma_p}{E_{sb}}.
    \label{heatfluxeq}
\end{equation}
The numerator is the heat flux, which is composed of the energy associated with the ion temperature $T_i$ and the energy from the electrode sheath potential drop $V_{sheath}$, multiplied by the particle flux $\Gamma_p$. The particle flux is calculated from the number density and the plasma sound speed, and the sheath potential drop is assumed to be the applied compression bank voltage, obtained from measurements on ZaP-HD. The heat flux is then divided by the surface binding energy $E_{sb}$ to obtain the sublimation flux. The value for $E_{sb}$ may be taken as the heat of sublimation, which is 7.4~eV for graphite. \cite{dasent1982inorganic} This assumes that all of the heat flux goes to heating the graphite. The sputtered flux $\Gamma_{sp}$ is calculated using
\begin{equation}
    \Gamma_{sp} = n_e c_s sin(\alpha) Y^*,
    \label{physsputeq}
\end{equation}
which assumes physical sputtering of carbon by $H^+$ ions at normal incidence ($\alpha=90^\circ$) to the surface and 100$\%$ effective sputtering yield $Y^*$. Both of these calculations represent the maximum possible flux under the stated assumptions.

Throughout most of the pulse duration, the erosion fluxes are at least two orders of magnitude larger than the maximum theoretical sputtered flux. The measured fluxes approach the maximum sublimation flux, and at times exceed it for the 9~kV pulses. This suggests that the majority of the C-III emission used for the S/XB measurement comes from ionized carbon neutrals that have sublimated off the electrode surface. When the current is low at early and late stages of the pulse, the measured erosion flux is at or below the theoretical sputtered flux limit. At these times, little to no radiation is measured on the spectrometer, therefore these values indicate the lower limit of the erosion measurement. Therefore, this data does not indicate whether sublimation occurs at these times at reduced fluxes. The duration of C-III emission also decreases with lower pinch current. The resulting decrease in plasma temperature and heat flux likely cause slower surface temperature heating and reduced ionization and excitation rates. A lower pinch current cannot sustain sublimation for as long, because the graphite is losing heat through radiation and conduction which the heat flux from the plasma and resistive heating must resupply for sublimation to continue. Notably, the measured erosion during peak current is far too large for physical sputtering to account for their production, leading to the hypothesis that sublimation is primarily responsible for the measured gross electrode erosion.

\subsection{\label{sec:exp2}Redeposition of sublimated particles reduces the net erosion}
% Analysis of eroded neutrale nergies, mean free paths, redeposition
The mechanism of material erosion implies a difference in the energy of the neutral that is ejected from the surface. Specifically, the energy of a sputtered atom should be comparable to the ion impact energy, and a sublimated atom should have an energy comparable to the electrode surface temperature. This difference in energy translates to a different ionization mean free path for the two neutral populations, which ultimately determines whether the neutral enters the bulk plasma or is redeposited on the electrode. The following describes the calculation of the ionization mean free path of neutral carbon from both erosion processes. In both cases, the neutral carbon enters the plasma at thermal speed $v_{th}$,
\begin{equation}
    v_{th,n} = \left( \frac{2k_BT_n}{m_n} \right)^{1/2},
    \label{thermalspeed}
\end{equation}
where $k_B$ is Boltzmann's constant, $T_n$ is the neutral temperature, and $m_n$ is the neutral mass. The mean free path for electron impact ionization of neutrals is given by
\begin{equation}
    \lambda_n = \frac{v_{th,n}}{\nu_n} = \frac{v_{th,n}}{n_e \overline{\sigma v}},
    \label{mfpeq}
\end{equation}
which is the ratio of the thermal speed and the collision frequency between electrons and neutrals, $\nu_n$. The collision frequency itself is the product of the electron number density $n_e$ and the ionization rate coefficient, $\overline{\sigma v}$. The ionization rate coefficients are  obtained by integrating the cross-section over a Maxwellian distribution of the electron velocity. Values of this coefficient from ADAS \cite{Summers_2011_ADAS} for single ionization of neutral carbon range from 10$^{-19}-10^{-13}$~m$^3$s$^{-1}$ over electron temperatures between 1 and 10$^4$~eV.

The next step is to determine the appropriate value of $T_n$ for sublimation and sputtering. Sublimated neutrals should have energy corresponding to the surface temperature. This gives a minimum $T_n$ of 3900~K, the sublimation temperature of graphite, \cite{ABRAHAMSON1974111} equivalent to 0.3~eV. 

In the case of sputtered neutrals, the neutral energy corresponds to the difference between the ion impact energy and the threshold energy for sputtering. Using the analysis presented in Ref.~\onlinecite{Stangeby_2000}, this energy is
\begin{equation}
    T_{n,sp} = E_0 \gamma(1 - \gamma) - E_{sb},
    \label{ejectedenergyeq}
\end{equation}
where $T_{n,sp}$ is the energy of the sputtered neutral, $E_0$ is the impact energy of the ion, $E_{sb}$ is the surface binding energy of the solid, and $\gamma$ is the maximum energy fraction defined as
\begin{equation}
    \gamma = \frac{4M_1 M_2}{(M_1 + M_2)^2}.
    \label{energyfrac}
\end{equation}
This fraction comes from the momentum transfer of two particles with mass $M_1$ and $M_2$ in a head-on collision. For the pure hydrogen plasma in ZaP-HD, $H^+$ is the dominant ion species. Therefore, the energy fraction is calculated with the atomic mass of hydrogen and carbon. The ion impact energy is calculated with
\begin{equation}
    E_0 = 2k_BT_i + \mid eV_{sheath} \mid.
    \label{impactenergy}
\end{equation}
$V_{sheath}$ is the voltage drop through the electrode sheath, assumed to be the applied voltage. This is approximated from voltage measurements to be between 2.8 and 4.8 kV. Assuming thermal equilibrium, $T_i = T_e = 1$~keV from Ref.~\onlinecite{Shumlak2017}. Calculating the impact energy with Eq.~\ref{impactenergy} gives $E_0=6.8$~keV, which corresponds to $T_{n,sp}=1.4$~keV from Eq.~\ref{ejectedenergyeq}. Note that $E_{sb}=7.4$~eV is negligible compared to the first term in Eq.~\ref{ejectedenergyeq} for the conditions being studied. The ionization mean free path of neutrals is then calculated with Eqs.~\ref{thermalspeed} and \ref{mfpeq}. The scale length of interest for comparison is the sheath thickness, $\lambda_s$, which is on the order of the Debye length $\lambda_D$,
\begin{equation}
    \lambda_s = \left( \frac{eV_{sheath}}{k_BT_e}\right)^{1/2}\lambda_D,
    \label{sheaththickness}
\end{equation}
using the electron number density $n_e=2\times10^{23}$~m$^{-3}$ from Ref.~\onlinecite{Shumlak2017}.

\begin{figure}
\includegraphics[scale=0.57]{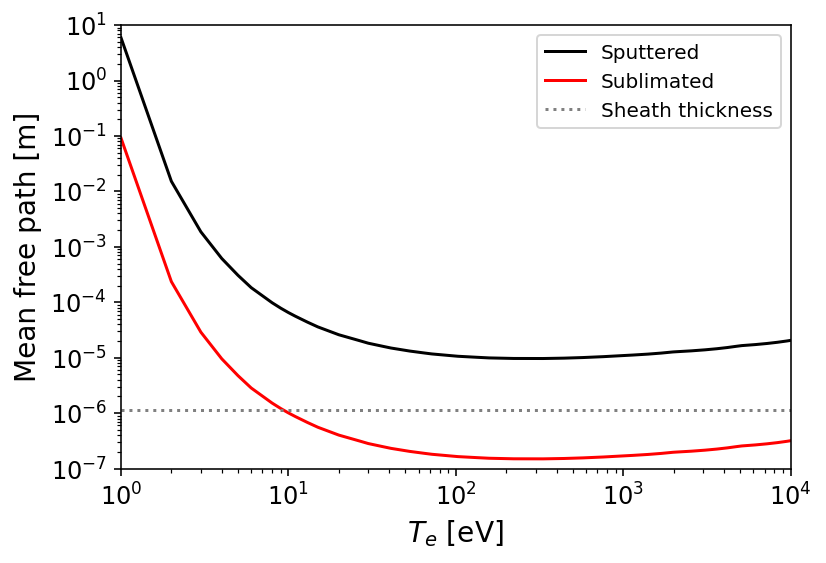}
\caption{\label{fig:mfpplot} Comparison of the electron impact ionization mean free paths for sputtered and sublimated carbon neutrals. For the $100-1000$~eV range of temperatures relevant to ZaP-HD, the sublimated neutrals are ionized over a distance much shorter than the sheath thickness, while sputtered neutrals undergo ionization well outside of the sheath.}
\end{figure}

The resulting mean free paths are plotted in Fig.~\ref{fig:mfpplot}. For the $100-1000$~eV range of electron temperatures expected on ZaP-HD, the sputtered neutrals have a mean free path of $1\times10^{-5}$~m, while sublimated neutrals have a mean free path of $2\times10^{-7}$~m. These values are compared to the sheath thickness calculated from Eq.~\ref{sheaththickness} which is $1\times10^{-6}$~m. This analysis shows that sublimated neutrals are ionized within the sheath, while sputtered neutrals are ionized over distances greater than the sheath thickness. The sublimated neutrals that are ionized within the sheath are subsequently accelerated by the sheath electric field back to the electrode surface where they are redeposited. By contrast, since sputtered neutrals are ionized much deeper within the bulk plasma they are not affected by the sheath electric field. Crucially, these results describe a physical process of recycling of eroded carbon atoms that significantly reduces the net erosion. Although sublimation accounts for the bulk of the total erosion, high redeposition rates return this carbon to the electrode, while the sputtered carbon flux is primarily responsible for the net erosion.

\subsection{Ex-situ analysis of electrode coupons}
\label{sec:exsitu}
Removable coupons permitted mass measurements and ex-situ analysis of the microscopic graphite surface after plasma exposure. The coupon mass was recorded before and after each experimental campaign using a Mettler Toledo MS403TS scale with a 1~mg accuracy. The average initial mass of the three coupons was 71.056~g, and each coupon experienced net mass loss.

To compare these mass measurements to the erosion measurements from Sec.~\ref{sec:exp1}, an equivalent net erosion flux, $\Gamma_{net}$, can be calculated using
\begin{equation}
    \Gamma_{net}  = \frac{\Delta m}{N_p m_C \tau_p A_p},
    \label{neterosionflux}
\end{equation}
where $\Delta m$ is the change in mass, $N_p$ is the number of pulses, $m_C$ is the carbon atomic mass, $\tau_p$ is the duration of plasma exposure, and $A_p$ is the area over which erosion occurs. Simplifying assumptions are made, namely that the erosion flux is constant throughout the duration of plasma exposure and over the entire specified area. The duration $\tau_p$ is defined for when the pinch current is above $20\%$ of the peak current using measurements from Fig.~\ref{fig:sxb_currents}(a), which is approximately 80~$\mu$s at each voltage setting. The area $A_p$ is assumed to be the cross-section of the pinch with the characteristic radius of 3~mm from Ref.~\onlinecite{Shumlak2017}.

Some additional quantities are required to calculate the electrode erosion rate and characterize plasma exposure. The net mass loss measurement is divided by the number of pulses to obtain the mass loss per pulse. This also assumes that the erosion rate is constant between pulses. The total charge passed through the electrode during an individual pulse is calculated by integrating the current traces in Fig.~\ref{fig:sxb_currents}(a). The ratio of these values gives the electrode erosion rate in mg/C units. The peak current density is calculated by dividing the peak pinch current from each current trace in Fig.~\ref{fig:sxb_currents}(a) by the area $A_p$ used in Eq.~\ref{neterosionflux}. In addition, the hourly mass loss rate is calculated in the same manner as Ref.~\onlinecite{Thompson_2023_POP}, which assumes the 10~Hz repetition rate of a power plant. The ZaP-HD transferred charge per pulse is multiplied by 10~Hz, which gives the charge transferred per second. Converting this to the charge transferred per hour and multiplying by the ZaP-HD erosion rates gives the hourly mass loss rate.

Lastly, the total ion fluence is calculated to quantify the overall plasma exposure on each coupon. Using the raw spectra of C-III emission recorded for the S/XB measurements in Fig.~\ref{fig:sxb_currents}(b), the ion temperature is extracted from Doppler broadened profiles using the basic method described in Ref.~\onlinecite{Forbes_2020_RSI}. Like the analysis in Sec.~\ref{sec:exp2}, thermal equilibrium between ions and electrons and the peak density from Ref.~\onlinecite{Shumlak2017} are assumed. The hydrogen ion flux is calculated with
\begin{equation}
    \Gamma_{H} = nc_{s} \approx \frac{1}{2}n_0[k(T_e + T_i)/m_H]^{1/2}, 
    \label{partfluxeq}
\end{equation}
where $m_H$ is the mass of the hydrogen ion. Integration of this particle flux over the duration of the measurement provides the particle fluence for each pulse, which is multiplied by the pulse count to get the total particle fluence.

The results of these calculations are summarized in Table~\ref{tab:coups}. Calculating the net erosion fluxes with Eq.~\ref{neterosionflux} gives values between $10^{27} - 10^{28}$~m$^{-2}$s$^{-1}$. These values are significantly lower than the measured total erosion fluxes during peak pinch current given in Fig.~\ref{fig:sxb_currents}(b). The net erosion fluxes are also lower than the theoretical sputtered flux by 1-2 orders of magnitude, but this theoretical value assumes a 100$\%$ sputtering yield made in Ref.~\onlinecite{Khairi_2025_RSI}. This result supports the assertion made in Sec.~\ref{sec:exp2} that net erosion is primarily due to sputtering.
\begin{table*}
\caption{\label{tab:coups}Summary of mass measurements, erosion-related quantities and exposure-related quantities for each graphite electrode coupon. Total fluence values refer to the entire experimental campaign for each plasma exposure case.}
\begin{ruledtabular}
\begin{tabular}{c c c c c c c c c}
 Case &
 \makecell[c]{Measured mass change, \\ $\Delta m$ \text{[mg]}} &
 \makecell[c]{Mass change \\ per pulse \text{[mg]}} &
 \makecell[c]{Charge transferred \\ per pulse \text{[C]}} &
 \makecell[c]{Net erosion flux \\ per pulse, $\Gamma_{net}$ \\ \text{[$\times$10$^{27}$ m$^{-2}$s$^{-1}$]}} &
 \makecell[c]{Erosion rate \\ \text{[mg/C]}} & 
 \makecell[c]{Current density \\ \text{[$\times$10$^9$A/m$^2$]}} &
 \makecell[c]{Mass loss \\ per hour [kg]} &
 \makecell[c]{Total ion fluence \\ \text{[$\times$10$^{26}$ m$^{-2}$]}} \\
 \hline
I   & $-17$ & $-0.085$ & 7.5 & $1.8$ & 0.01 & $4.8$ & 0.003 & $4.4$\\
II  & $-32$ & $-0.64$  & 6.2 & $13.1$ & 0.10 & $4.0$ & 0.02 & $1.0$\\
III & $-4$  & $-0.095$ & 5.0 & $2.0$ & 0.02 & $3.5$ & 0.003 & $0.5$\\
\end{tabular}
\end{ruledtabular}
\end{table*}

The erosion rates of the electrode coupons are between 0.01 and 0.1~mg/C. The lower end of this range is comparable to 0.016~mg/C for a carbon electrode operating in a 100~A vacuum arc discharge, \cite{Kimblin_1974} while the upper end remains below the 0.250~mg/C value for a simulated 80~kA carbon arc furnace. \cite{Sævarsdóttir_2016} The erosion rates on ZaP-HD are also significantly lower than the 8~mg/C observed for a tungsten electrode in a 120~kA high-pressure discharge-current. \cite{Bogomaz2003} The current densities are within the range defined for cathodic arc discharges of $10^8 - 10^{13}$~A/m$^2$ given in Ref.~\onlinecite{Anders_2024}.

Notably, net erosion does not appear to increase monotonically with the intensity of plasma exposure, since the highest erosion rate was observed for the intermediate conditions of Case II. This result may be explained by the presence of competing effects that contribute to net erosion. The analysis in Sec.~\ref{sec:exp2} describes how sputtered carbon is primarily responsible for the net erosion. An increase in pinch current implies increased plasma temperature and density from the scaling relations described in Sec.~\ref{sec:zaphd}. The pinch current is increased through a higher capacitor bank voltage, which applies a larger bias voltage to the electrode. These factors increase the ion impact energy, which is shown to be on the order of several keV in Sec.~\ref{sec:exp2}. At these impact energies, sputtering yields begin to decrease due to the ions penetrating deeper into the solid lattice. This reduced physical sputtering yield could result in decreased net erosion despite the increased plasma parameters.

The hourly mass loss rates calculated using the erosion rates from ZaP-HD are significantly lower than the $\approx3$~kg per hour estimated in Ref.~\onlinecite{Thompson_2023_POP}, which assumes the 0.250~mg/C erosion rate from Ref.~\onlinecite{Sævarsdóttir_2016}. While this result is quite favorable, a power plant level device will operate at pinch currents over a thousand times larger than ZaP-HD and will likely cause significantly greater erosion rates. Finally, total ion fluences are 1-2 orders of magnitude lower than for LPDs. \cite{Linsmeier_2017_Facilities} This limits the evaluation of PMI effects at comparably high particle fluence and long plasma exposures, albeit at the elevated particle and heat fluxes on ZaP-HD.

\subsubsection{Electron microscopy}
 
Imaging of the microscopic graphite surface can reveal changes in the surface morphology that are indicative of certain PMI processes. For instance, surface modification in the form of pits or craters may indicate erosion by sputtering processes. Layers of material deposited on the surface can result in distinctive textures or features that are characteristic of certain processes.

\begin{figure*}
\includegraphics[scale=0.59]{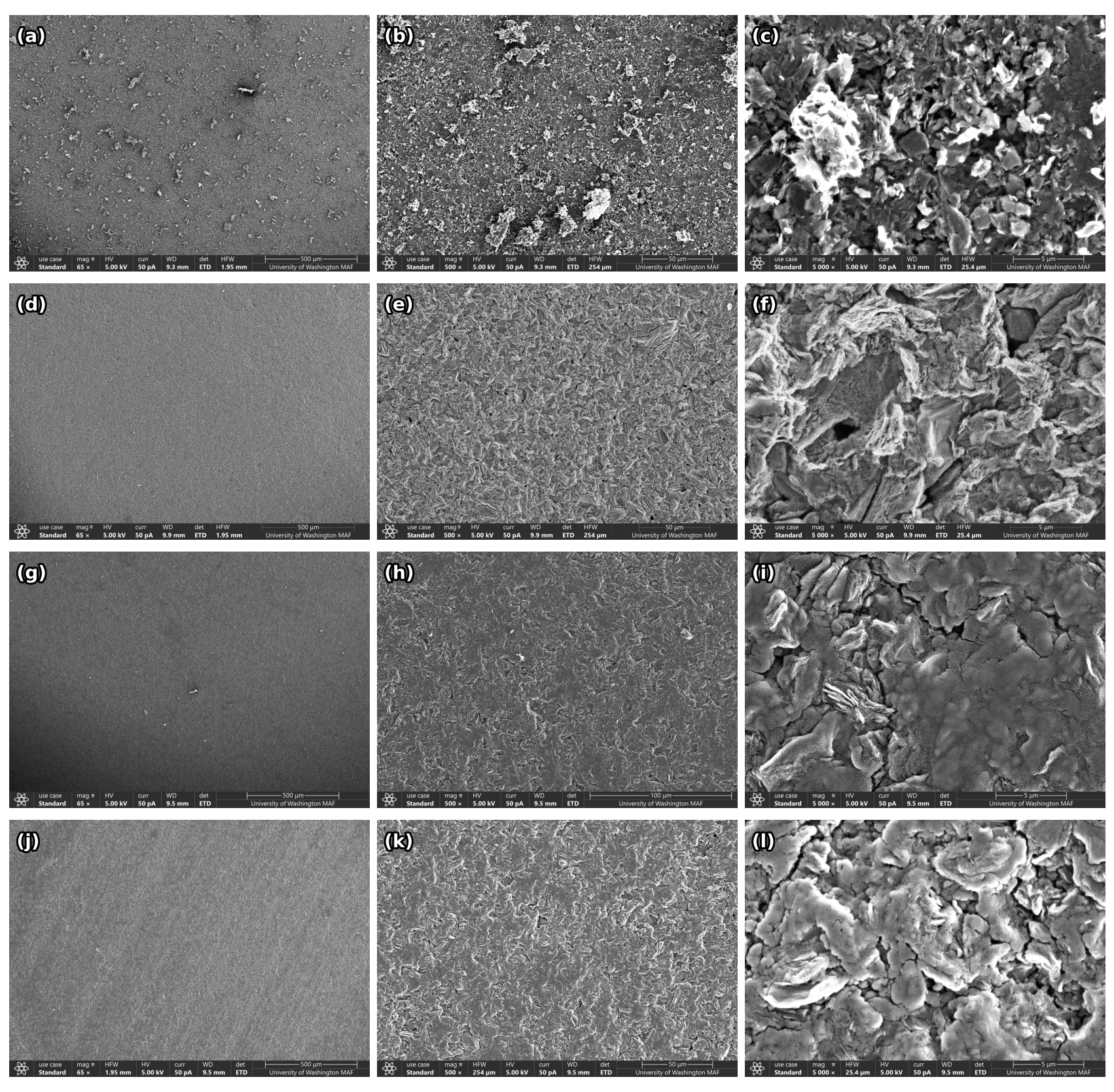}
\caption{SEM micrographs of the graphite electrode coupon surface before plasma exposure (a), (b), (c), and after plasma exposure for Case I (d), (e), (f), Case II (g), (h), (i), and Case III (j), (k), (l). The intensity of plasma exposure on the coupon decreases from the second row downwards. From left to right, the magnification for each column is 65x, 500x, and 5000x. All plasma exposures resulted in removal of large particles scattered over the surface, and formation of irregular, consolidated features.}
\label{fig:sem_all}
\end{figure*}

Micrographs of the electrode coupon surface were obtained with the Apreo 2 SEM using the secondary electron detector. Secondary electron emission occurs in the topmost layers of the material, and therefore is more indicative of surface details. A set of micrographs is provided in Fig.~\ref{fig:sem_all}. Magnification increases left to right, and plasma exposure decreases from the second row downwards. The surface prior to exposure, shown in Figs.~\ref{fig:sem_all}(a)-(c), consists of distinct particles or clumps of particles 5-100~$\mu$m in size that are scattered over the bulk graphite matrix. These particles are larger than the specified particle size for this graphite grade, approximately 5~$\mu$m, \cite{POCO} and were likely produced by fracturing of the matrix during machining. At the 5000x magnification, Fig.~\ref{fig:sem_all}(c), the granular structure of the matrix is more apparent. Some porosity at the surface is observable, indicated by dark voids in between grains. These voids are consistent with the specified pore size of 0.8 $\mu$m. \cite{POCO}

After plasma exposure, there are clear changes to the surface morphology. The 65x magnification images, Figs.~\ref{fig:sem_all}(d), (g), and (j), show that the large particles are almost completely removed. Increasing the magnification to 500x in Figs.~\ref{fig:sem_all}(e), (h) and (k) reveals consolidated surface structures that resemble flakes, scales, or ripples. At the 5000x magnification, greater detail of these structures is observed. Their shapes are irregular, ranging from ribbon-like striations in Fig.~\ref{fig:sem_all}(f) to more rounded slabs in Figs.~\ref{fig:sem_all}(i) and (l). Some areas, such as the darker region in the right-central portion of Fig.~\ref{fig:sem_all}(i) appear to be relatively smooth. Lastly, cracking of the surface is observed at 500x and 5000x magnification, which may be an indication of thermal fatigue from many heating cycles. This suggests a degree of influence of the surface temperature in creating this morphology.

Quantitative analysis of SEM micrographs is performed along the $x$ and $y$ axes established in Fig.~\ref{fig:surfmap}. Two quantities are calculated: the first is the average area of dark regions which indicates the presence of voids in the graphite matrix, and the second is the average eccentricity of these dark regions which characterizes their change in shape. A Python script converts raw micrograph images to grayscale and applies a 15$\%$ intensity threshold to isolate dark pixels. A filter is applied to remove detected regions smaller than 0.5~microns. The result is detection of the qualitative features described as voids, gaps, and cracks. For each contiguous region detected, the area and the eccentricity are calculated. 

\begin{figure}
     \begin{subfigure}[b]{0.4\textwidth}
         \centering
         \includegraphics[width=\textwidth]{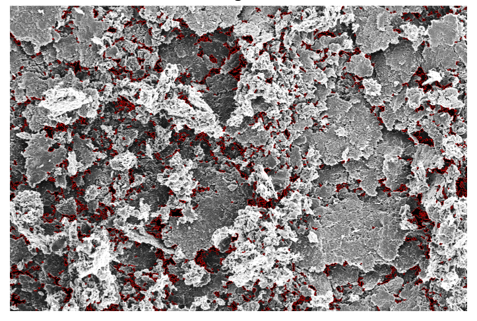}
         \caption{Before plasma exposure}
         \label{fig:darkdetpre}
     \end{subfigure}
     \hspace{0.01cm} % Adjust horizontal spacing here
     \begin{subfigure}[b]{0.4\textwidth}
         \centering
         \includegraphics[width=\textwidth]{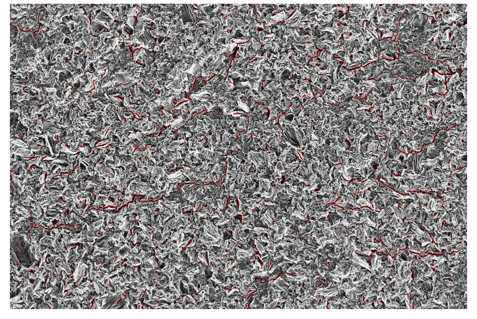}
         \caption{After plasma exposure}
         \label{fig:darkdetpost}
     \end{subfigure}
        \caption{Dark region detection in a sample SEM micrograph of the graphite coupon surface before and after plasma exposure. These images were taken for the coupon in Case I at 500x magnification. In (a), the detected dark regions are the irregularly shaped voids in between graphite grains. In (b), the detected dark regions are the cracks and narrow gaps in between consolidated ribbon-like structures.}
        \label{fig:darkdet}
\end{figure}

\begin{figure*}
     \centering
     \begin{subfigure}[b]{0.4\textwidth}
         \centering
         \includegraphics[width=\textwidth]{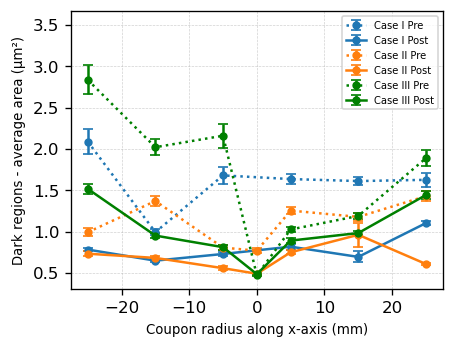}
         \caption{}
         \label{fig:darkx}
     \end{subfigure}
     \hspace{0.05cm} % Adjust horizontal spacing here
     \begin{subfigure}[b]{0.4\textwidth}
         \centering
         \includegraphics[width=\textwidth]{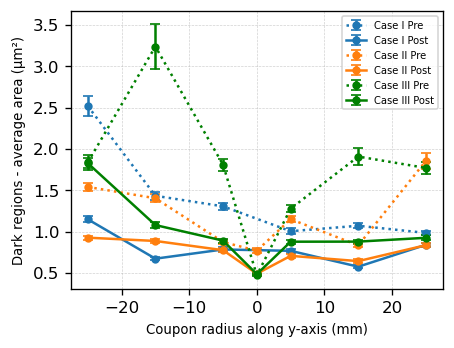}
         \caption{}
         \label{fig:darky}
     \end{subfigure}
        \caption{Changes in the average area of dark regions in SEM micrographs of the graphite coupon surface along the (a) $x$ axis and  (b) $y$ axis defined in Fig.~\ref{fig:surfmap}. Analysis is performed for micrographs at 500x magnification. For all three plasma exposure cases, the dark region area decreases along both axes. This represents the change in morphology of the surface from the porous graphite matrix to the irregular, consolidated structures observed in Fig.~\ref{fig:sem_all}.}
        \label{fig:darkarea}
\end{figure*}

\begin{figure*}
     \centering
     \begin{subfigure}[b]{0.4\textwidth}
         \centering
         \includegraphics[width=\textwidth]{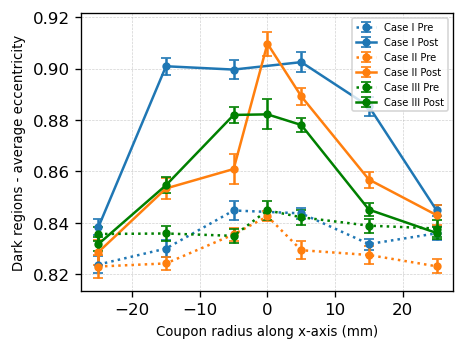}
         \caption{}
         \label{fig:eccx}
     \end{subfigure}
     \hspace{0.05cm} % Adjust horizontal spacing here
     \begin{subfigure}[b]{0.4\textwidth}
         \centering
         \includegraphics[width=\textwidth]{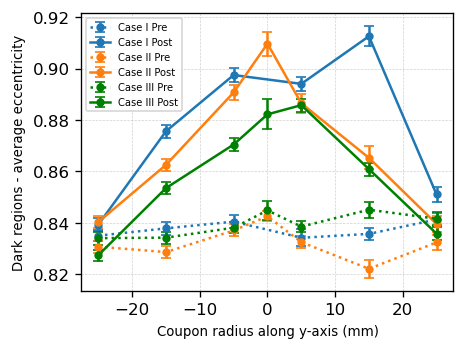}
         \caption{}
         \label{fig:eccy}
     \end{subfigure}
        \caption{Changes in the average eccentricity of the detected dark regions in SEM micrographs of the graphite coupon surface along the (a) $x$ axis and (b) $y$ axis defined in Fig.~\ref{fig:surfmap}. Analysis is performed for micrographs at 500x magnification. For all three plasma exposure cases, the eccentricity increases significantly, representing a change in morphology from the voids in the porous graphite matrix to the narrow gaps and cracks in and around the irregular, consolidated structures observed in Fig.~\ref{fig:sem_all}.}
        \label{fig:darkecc}
\end{figure*}

An example of the detected dark regions at 500x magnification is given in Fig.~\ref{fig:darkdet}. Before plasma exposure, in Fig.~\ref{fig:darkdet}(a), the dark regions are the voids within the graphite matrix. After plasma exposure, in Fig.~\ref{fig:darkdet}(b), the detected dark regions are primarily long, narrow gaps and cracks in between the consolidated ribbon-like structures. The spatial profiles of the average area and average eccentricity of dark regions detected at 500x magnification are plotted in Figs.~\ref{fig:darkarea} and \ref{fig:darkecc} respectively. Error bars represent the standard deviation of the area or eccentricity of detected regions within a micrograph divided by the square root of the number of detected regions. It is evident that this change in morphology occurs throughout the 50~mm span of the SEM measurements, in perpendicular directions, and for all plasma exposure cases. This is a considerable portion of the coupon surface, which has a roughly 70~mm diameter as shown in Fig.~\ref{fig:surfmap}. Notably, this effect is observed over areas much larger than the reported ZaP-HD Z-pinch radius of $2-5$~mm. \cite{Ross2016RSI, Shumlak2017} This may be explained by wandering or expansion of the pinch column as it makes contact with the electrode, resulting in particle and heat flux over an extended area.

The observed morphological changes indicate some form of material rearrangement, which is consistent with the general process of carbon recycling posited in Sections~\ref{sec:exp1} and \ref{sec:exp2}. Attributing morphology changes to specific mechanisms such as sputtering or sublimation is not unequivocal, especially with possible redeposited layers obscuring initial surface modifications. However, some useful statements can still be made. For instance, there is a distinct lack of cavities or pitting of the surface that has been attributed to sputtering in graphite bombarded with energetic ions. \cite{SEYEDHABASHI_2025} Sputtering processes likely eject not only individual atoms but clusters of atoms \cite{Oyarzabal_2008} and even hydrocarbons. \cite{DAVIS1988234} If sputtering were more dominant, similar morphology would be expected on the graphite coupons. In terms of carbon deposition, the observed morphology is also unlike that of the cathode deposits formed by carbon arcs for nanosynthesis \cite{NG_2014}, which feature highly spherical morphology for surface temperatures exceeding 3000~K, and ion temperatures of about 1~eV. Due to the short ionization mean free path of sublimated neutrals calculated in Sec.~\ref{sec:exp2}, the redeposited layer on the graphite coupons is not expected to display such a preferential morphology, but would likely be more uniform or follow the contours of the surface.

While the discussion of graphite thermal effects has so far only addressed sublimation, previous work identified the possibility of graphite melting. \cite{Forbes_2020_PhD} This can occur at 4000~K for sufficiently high pressures. However, a large degree of uncertainty applies to the necessary pressures, which range from 110~atm and 100,000~atm. \cite{Bundy_1980} Calculation of the required plasma temperature using Eq.~\ref{radeq} to achieve these pressures gives a range of 320~eV to 320~keV, assuming $n_e = 2 \times10^{23}$~m$^{-3}$. Measured ion and electron temperatures on ZaP-HD either meet or exceed the low end of this range, \cite{Forbes_2020_RSI, Shumlak2017} lending some plausibility to graphite melting. The solidification of a melt layer on the surface of the graphite could explain the consolidated features observed in Fig.~\ref{fig:sem_all}.

\subsubsection{Profilometry}

Profilometry measurements provide a three-dimensional picture of the surface topography. Although a sense of depth can be perceived by the contrast in certain SEM measurements, these images do not explicitly capture height information. Measuring the surface profile can be used to determine the extent of material deposition or erosion, and quantifying the surface roughness has important implications on PMI effects such as sputtering, where the local angle of incidence can significantly impact the yield.

\begin{figure*}
     \centering
     \begin{subfigure}[b]{0.27\textwidth}
         \centering
         \includegraphics[width=\textwidth]{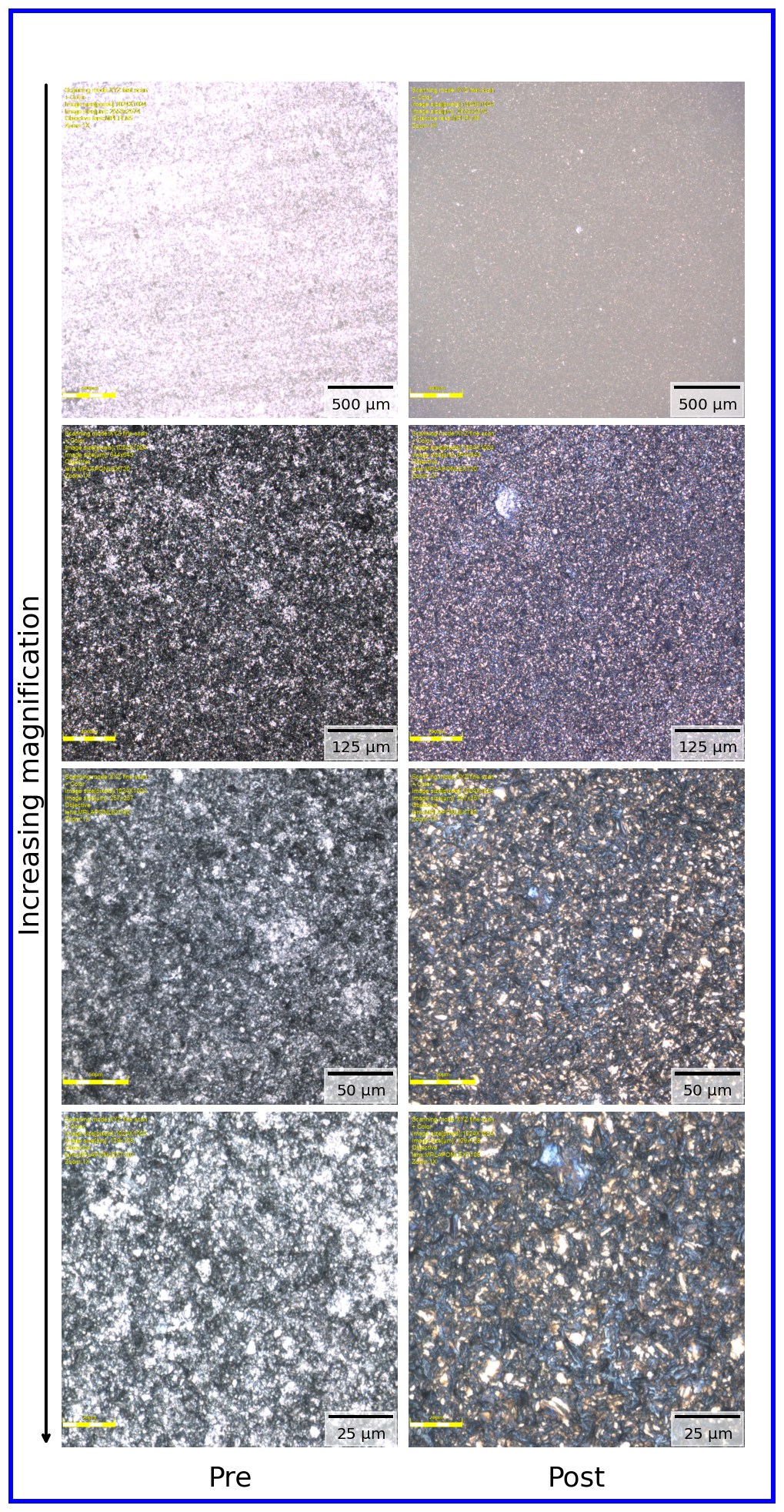}
         \caption{Case I}
         \label{fig:opticalI}
     \end{subfigure}
     \hspace{0.01cm} % Adjust horizontal spacing here
     \begin{subfigure}[b]{0.27\textwidth}
         \centering
         \includegraphics[width=\textwidth]{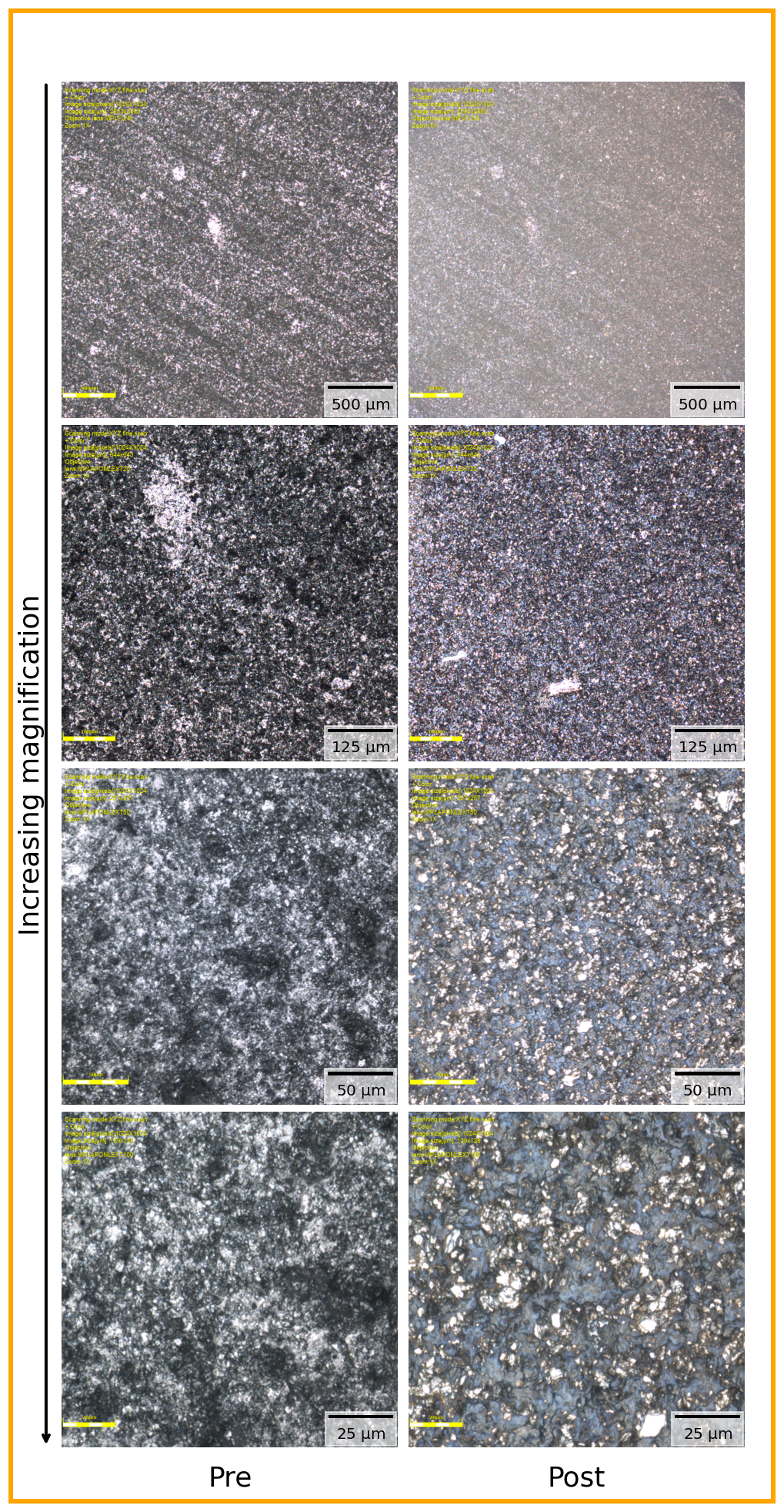}
         \caption{Case II}
         \label{fig:opticalII}
     \end{subfigure}
     \hspace{0.01cm} % Adjust horizontal spacing here
     \begin{subfigure}[b]{0.27\textwidth}
         \centering
         \includegraphics[width=\textwidth]{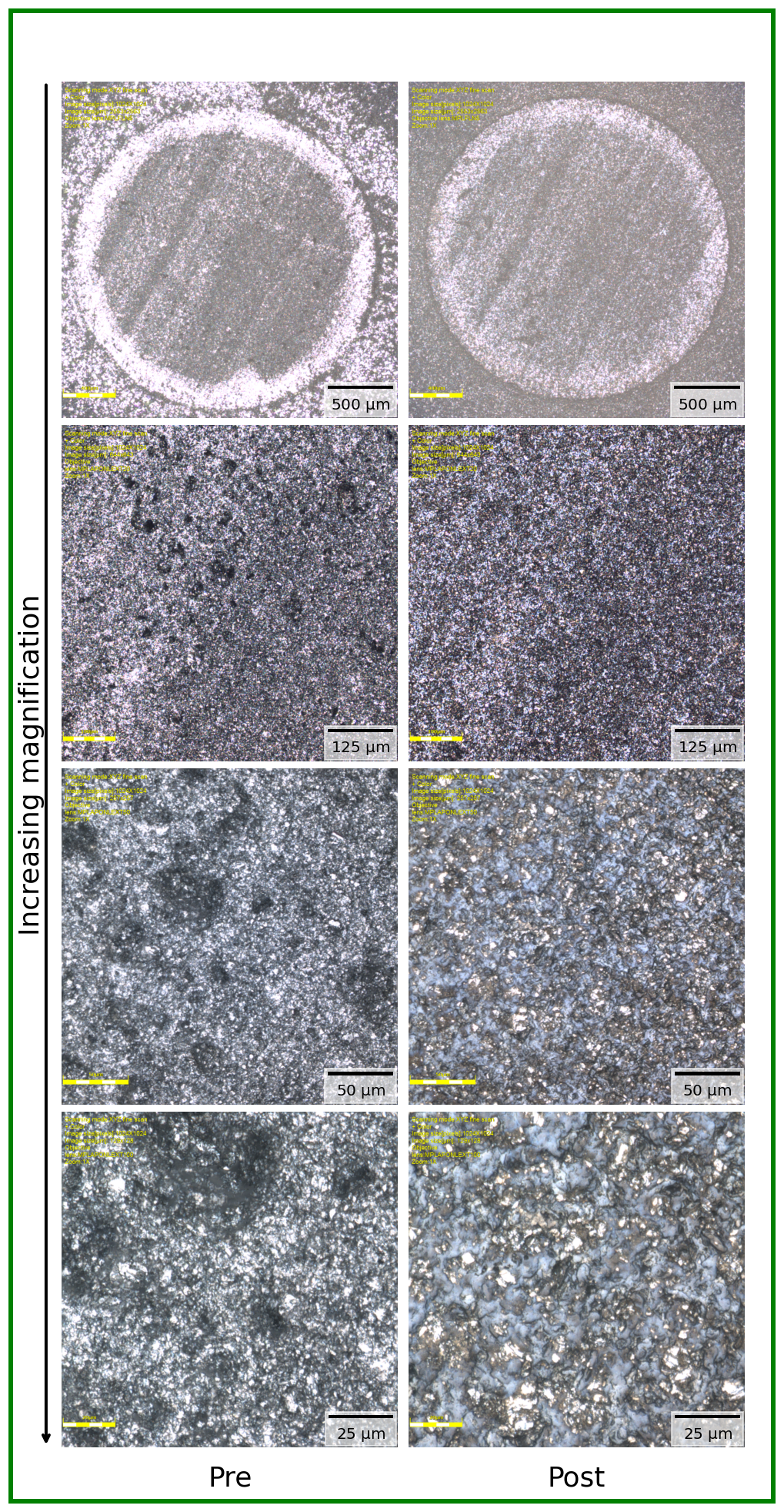}
         \caption{Case III}
         \label{fig:opticalIII}
     \end{subfigure}
        \caption{Optical images of the graphite coupon surface before and after plasma exposure for three plasma exposure cases. From top to bottom, the magnifications are 5x, 20x, 50x, and 100x. At lower magnifications, machining artifacts remain visible except for Case I which tested the highest plasma exposure conditions. Irregular, consolidated structures are observed at higher magnifications for all cases.}
        \label{fig:opticalpics}
\end{figure*}

\begin{figure*}
     \centering
     \begin{subfigure}[b]{0.27\textwidth}
         \centering
         \includegraphics[width=\textwidth]{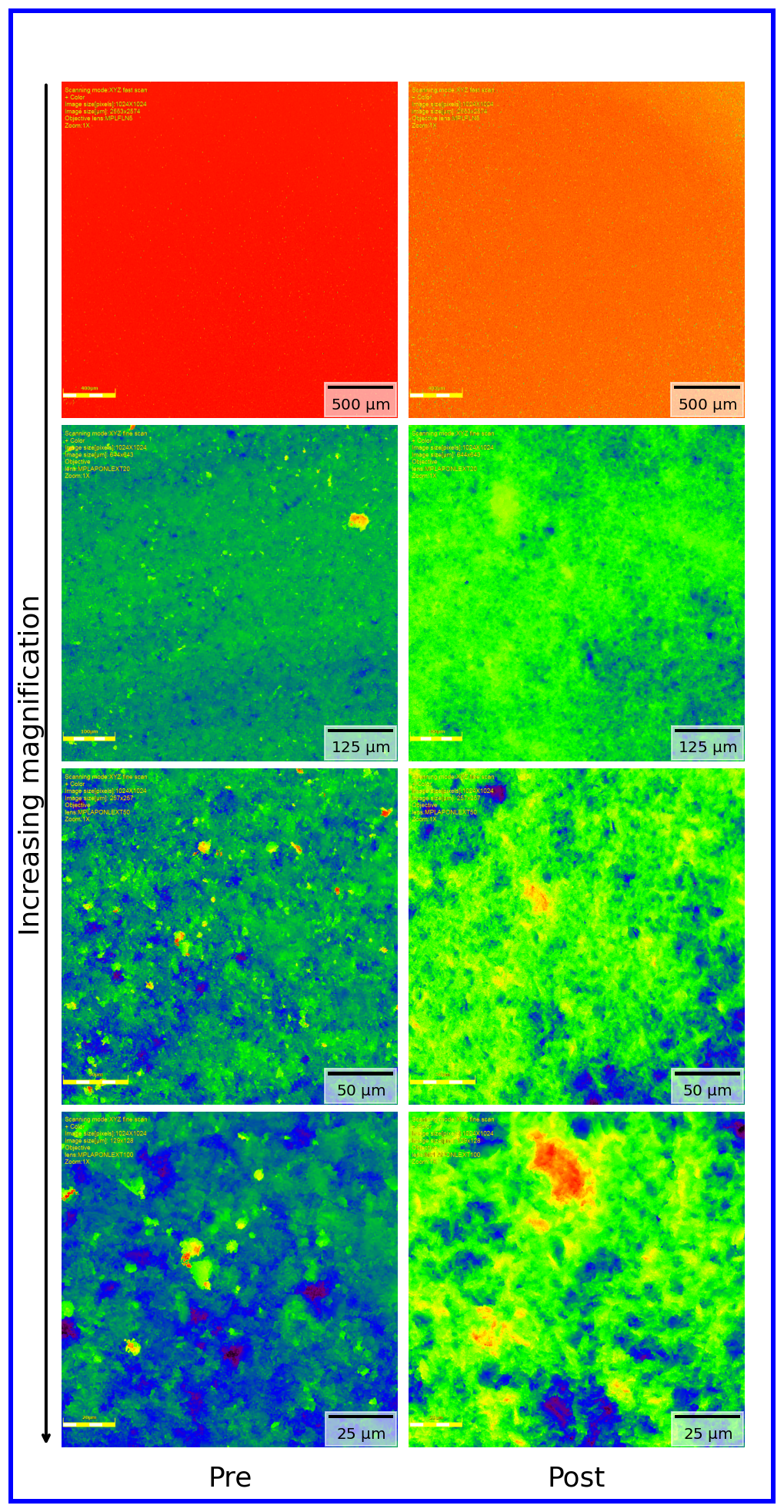}
         \caption{Case I}
         \label{fig:heightI}
     \end{subfigure}
     \hspace{0.01cm} % Adjust horizontal spacing here
     \begin{subfigure}[b]{0.27\textwidth}
         \centering
         \includegraphics[width=\textwidth]{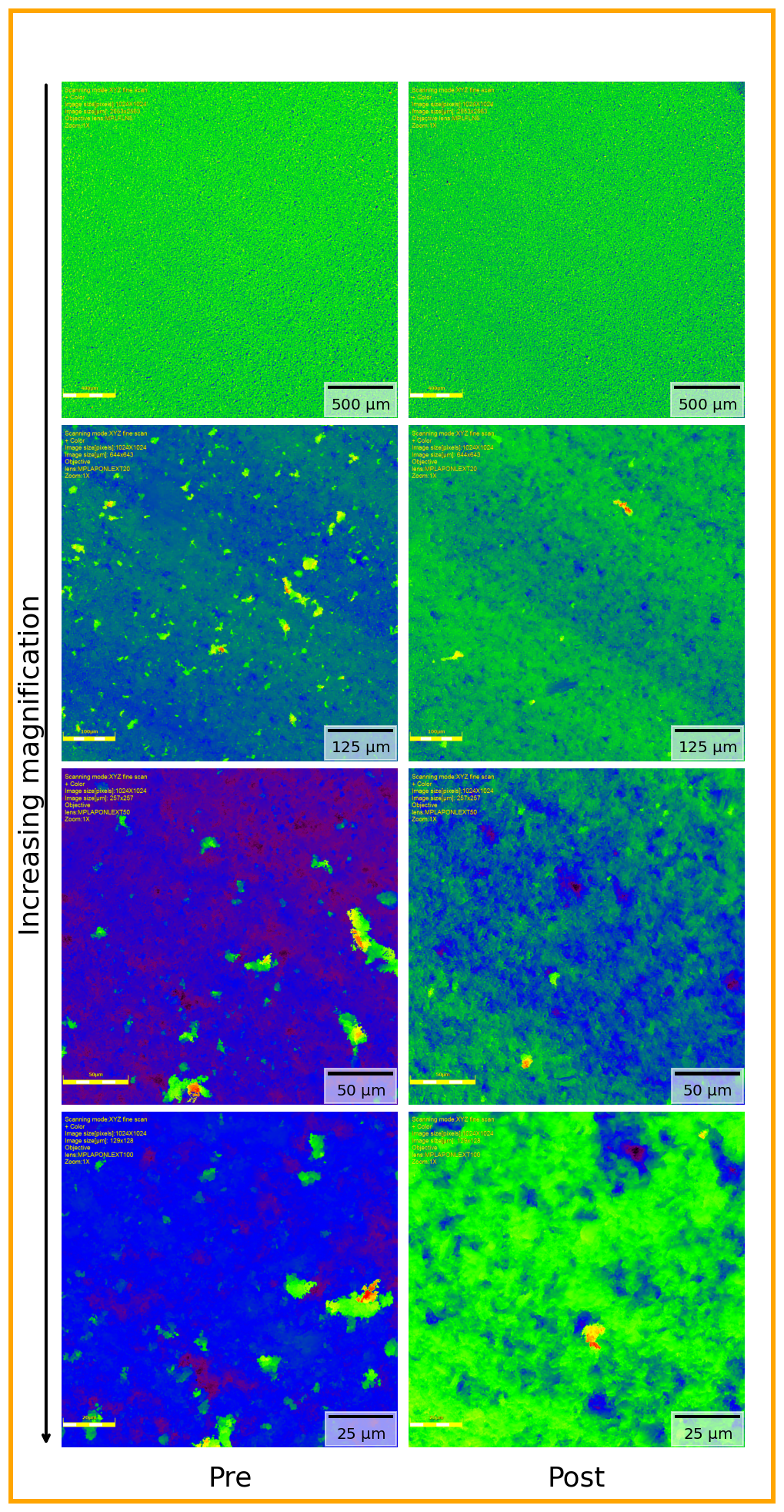}
         \caption{Case II}
         \label{fig:heightII}
     \end{subfigure}
     \hspace{0.01cm} % Adjust horizontal spacing here
     \begin{subfigure}[b]{0.27\textwidth}
         \centering
         \includegraphics[width=\textwidth]{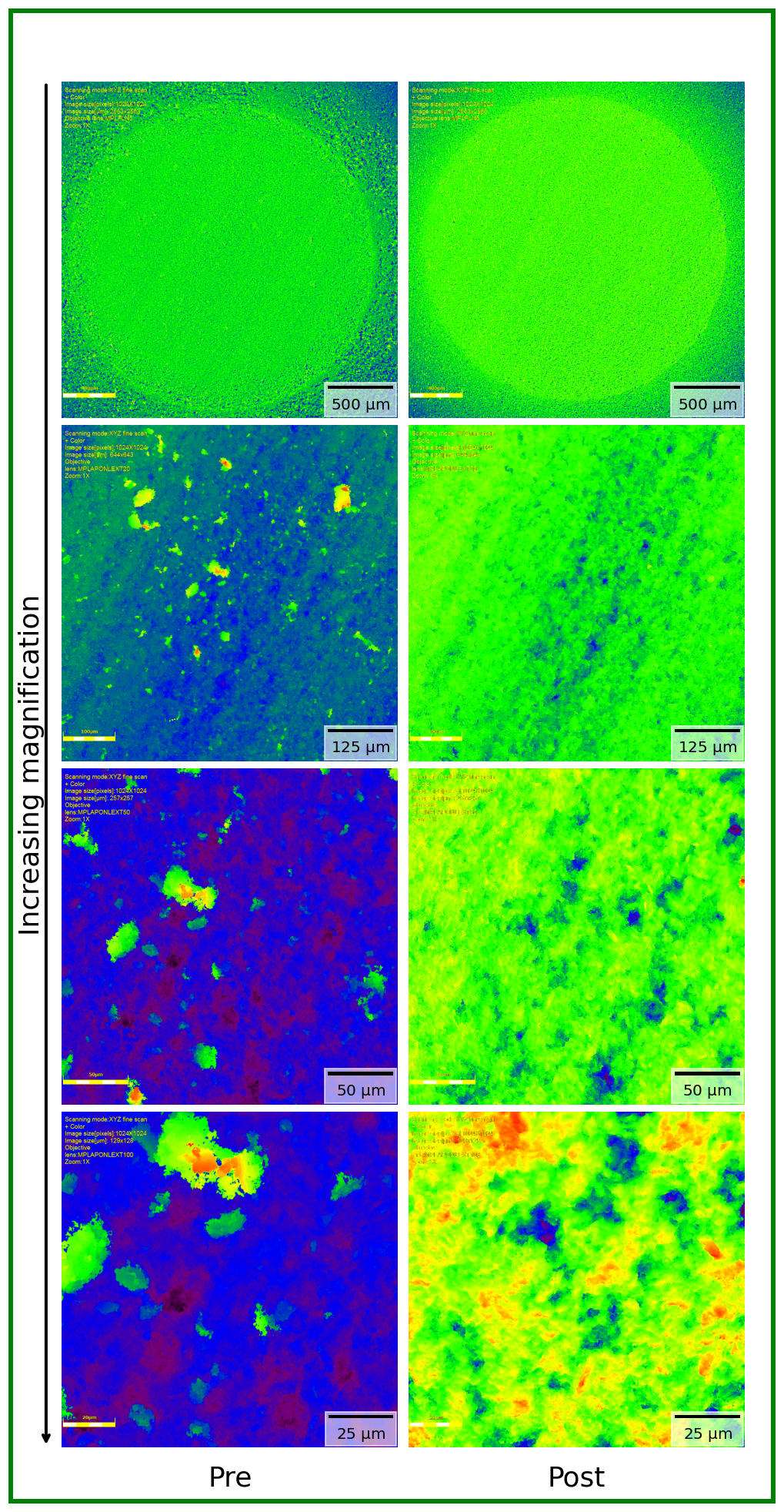}
         \caption{Case III}
         \label{fig:heightIII}
     \end{subfigure}
        \caption{Surface profile maps of the graphite coupon surface before and after plasma exposure for three plasma exposure cases. These maps directly correspond to the optical images in Fig.~\ref{fig:opticalpics}. From top to bottom, the magnifications are 5x, 20x, 50x, and 100x. As seen in SEM micrographs, removal of large particles is observed, along with the appearance of irregular, consolidated structures at higher magnification.}
        \label{fig:heightpics}
\end{figure*}

Profilometry measurements were taken with the Olympus OLS4100 optical profilometer. Optical images of the graphite surface and the corresponding height maps are given in Figs.~\ref{fig:opticalpics} and \ref{fig:heightpics} respectively. Each $4\times2$ grid of images contains measurements for a single coupon. For each grid, the left hand column contains images before plasma exposure, and the right hand column contains images after plasma exposure. Images were taken at 5x, 20x, 50x, and 100x magnifications, increasing down the rows. 

Before plasma exposure, the 5x magnification images in Fig.~\ref{fig:opticalpics} show parallel striations on the graphite surface left behind from the fabrication process. In Fig.~\ref{fig:opticalpics}(a) for Case I, the striations have been completely removed. The striations remain visible for Cases II and III, likely due to the lower ion fluence as listed in Table~\ref{tab:coups}. At 50x and 100x magnification, the effect of plasma exposure is qualitatively similar across all three cases, with the appearance of bright regions with rounded, smooth edges. These are likely the same features visible at 500x magnification in Fig.~\ref{fig:sem_all}. 

For consideration of the height maps in Fig.~\ref{fig:heightpics} there are two points of note. Firstly, the highly reflective surface in the top left image of Fig.~\ref{fig:opticalI} likely caused saturation of the sensor, which results in the flat height profile seen in the corresponding profile shown in Fig.~\ref{fig:heightI}. All other acquisitions were able to resolve the surface profile. Secondly, the reference point for zero height was not maintained between acquisitions, therefore absolute comparison of the heights measured is not possible. Consequently, only relative changes such as the roughness are calculated and compared. 

A summary of this analysis is provided in Tables~\ref{tab:profileresults5x} and \ref{tab:profileresults100x}, consisting of two roughness parameters, $R_a$ and $R_{RMS}$ at 5x and 100x magnifications. The parameter $R_a$ is the arithmetic average of the absolute deviations from the mean surface height, which captures the overall roughness but is less sensitive to large irregularities. The parameter $R_{RMS}$ is the root-mean-square (RMS) of the deviations from the mean surface height, which is more sensitive to irregularities. The values given are the average of the roughness parameter calculated along the vertical dimension of each image at 1024 sampling positions, i.e. at the pixel resolution of the profilometer height maps in Fig.~\ref{fig:heightpics}. The uncertainty provided is the standard deviation over these sampling locations. In general, both roughness parameters decrease after plasma exposure, and the variation across the sampled length also decreases. This is consistent with the removal of distinct particles and appearance of the consolidated features observed in SEM micrographs.
\begin{table}
\caption{\label{tab:profileresults5x}Summary of surface roughness measurements performed on the graphite electrode coupons before and after exposure to ZaP-HD Z-pinch plasmas. Profilometer measurements were made at 5x magnification. }
\begin{ruledtabular}
\setlength{\tabcolsep}{6pt} % increase horizontal spacing between columns
\begin{tabular}{c|cc|cc}
\multirow{2}{*}{Case $\#$} 
  & \multicolumn{2}{c|}{\underline{$R_a$ [$\mu$m]}} 
  & \multicolumn{2}{c}{\underline{$R_{RMS}$ [$\mu$m]}} \\
 & Before & After & Before & After \\
\hline
I  & 8.8 $\pm$ 1.2 & 18.1 $\pm$ 4.5 & 51.2 $\pm$ 4.4 & 53.4 $\pm$ 12.2 \\
II & 14.8 $\pm$ 0.9 & 13.3 $\pm$ 0.8  & 20.0 $\pm$ 1.3 & 17.7 $\pm$ 1.0  \\
III  & 19.0 $\pm$ 4.2 & 16.8 $\pm$ 1.9  & 26.4 $\pm$ 5.5 & 21.8 $\pm$ 2.3  \\
\end{tabular}
\end{ruledtabular}
\end{table}
\begin{table}
\caption{\label{tab:profileresults100x}Summary of surface roughness measurements performed on the graphite electrode coupons before and after exposure to ZaP-HD Z-pinch plasmas. Profilometer measurements were made at 100x magnification.}
\begin{ruledtabular}
\setlength{\tabcolsep}{6pt} % increase horizontal spacing between columns
\begin{tabular}{c|cc|cc}
\multirow{2}{*}{Case $\#$} 
  & \multicolumn{2}{c|}{\underline{$R_a$ [$\mu$m]}} 
  & \multicolumn{2}{c}{\underline{$R_{RMS}$ [$\mu$m]}} \\
 & Before & After & Before & After \\
\hline
I  & 0.6 $\pm$ 0.2 & 0.9 $\pm$ 0.3 & 0.8 $\pm$ 0.3 & 1.2 $\pm$ 0.3 \\
II & 0.8 $\pm$ 0.4 & 0.6 $\pm$ 0.2  & 1.3 $\pm$ 0.7 & 0.8 $\pm$ 0.2  \\
III  & 2.2$\pm$ 1.4 & 0.8 $\pm$ 0.2  & 3.1 $\pm$ 1.9 & 1.0 $\pm$ 0.2  \\
\end{tabular}
\end{ruledtabular}
\end{table}

\begin{figure}
\includegraphics[scale=0.5]{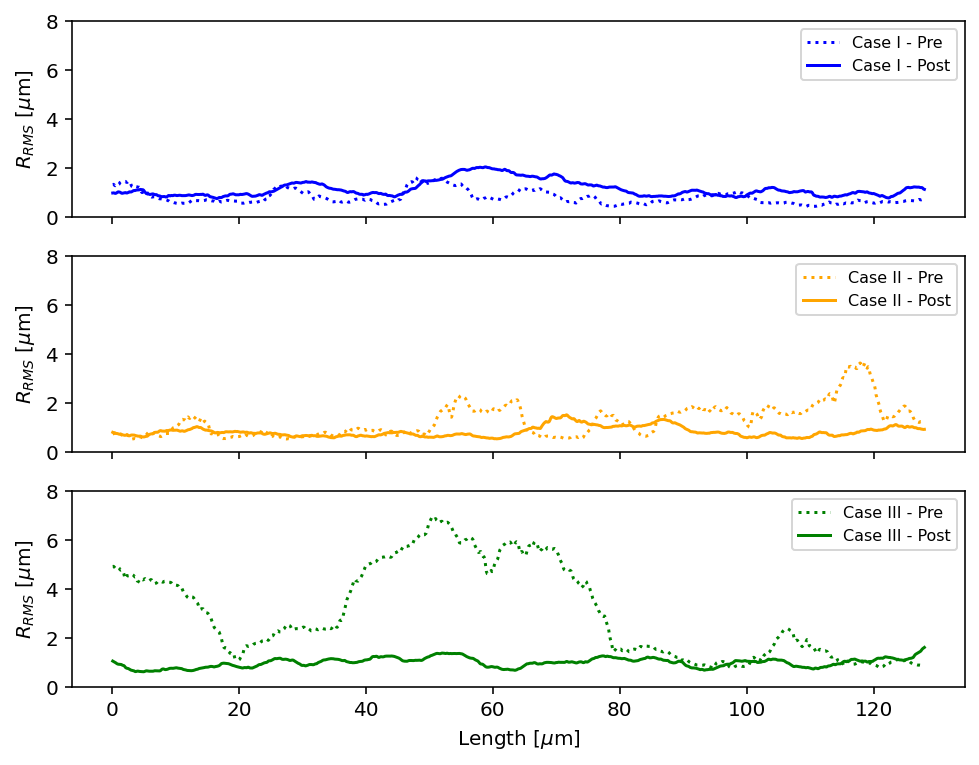}
\caption{\label{fig:roughnessprofiles}Spatial variation in the roughness parameter $R_{RMS}$ before and after three plasma exposure conditions. Smoothing of the surface occurs for the low fluence Case II and III, while slight roughening occurs at the high particle fluence of Case I.}
\end{figure}

The exception to this behavior occurs for Case I, which has an increase in roughness calculated for both 5x and 100x magnifications. Since similar morphological changes occur for all three cases, it is possible that the larger particle fluence in Case I was enough to induce surface changes beyond the initial smoothing effect. The large increase in roughness in $R_a$ for Case I in Table~\ref{tab:profileresults5x} is attributed to the high reflectivity observed for the 5x magnification image of Fig.~\ref{fig:opticalpics}(a), therefore is not indicative of the true change in roughness. Profiles of $R_{RMS}$ used to calculate the average values in Table~\ref{tab:profileresults100x} are plotted in Fig.~\ref{fig:roughnessprofiles} to illustrate the spatial variation of roughness. Again, smoothing of the surface occurs after plasma exposure with the exception of the coupon in Case I.

Smoothing of graphite electrode surfaces has been attributed to sublimation in certain conditions. A 30~kA spark gap in a nitrogen environment resulted in extremely smooth sub-micron scale surfaces. This was attributed to sublimation on the basis of a lack of craters that would indicate sputtering of graphite particles, as well as the appearance of deposits of monatomic layers of amorphous carbon. \cite{Gordon_1982} However, the effects of chemical reactions between carbon and nitrogen complicate a direct comparison. On the other hand, the importance of ion fluence and surface temperature has been noted in the submicron-scale roughening of graphite under argon ion bombardment, \cite{Eklund_1991} which aligns with the increased roughness observed for the graphite coupon in Case I.

\section{\label{sec:conc}Conclusion}
% Summary of main points
In this paper, the results of experiments studying the plasma-electrode interactions on the ZaP-HD SFS Z-pinch device are presented. In-situ measurements of the gross carbon erosion flux from the graphite electrode were conducted with S/XB spectroscopy. The measurements show agreement with the expected sublimation flux, but are significantly larger than the expected physical sputtering flux. This suggests that sublimation dominates the erosion process. However, analysis of the electron ionization mean free paths indicates that the relatively low energy neutrals produced by sublimation are ionized within the electrode sheath and redeposited back on the electrode. The sputtered neutrals have much higher energy and are ionized outside of the sheath. Sputtering is therefore primarily responsible for net erosion. This describes a physical mechanism for recycling of sublimated carbon and loss of sputtered carbon that results in significant reduction of net erosion.

Removable electrode coupons enabled mass measurements and ex-situ surface analysis for three plasma exposure conditions that varied the pinch current and the number of pulses. Net erosion fluxes calculated from mass-loss measurements are found to be significantly lower than spectroscopic measurements of the gross erosion, but are comparable to the expected sputtering flux. This result is consistent with the analysis on recycling of sublimated carbon. Net erosion rates in units of mg/C are comparable to or lower than those for electrodes on a variety of arc discharges. The largest erosion rate of 0.10 mg/C is observed for the intermediate plasma exposure condition, which suggests possible competing effects associated with increasing current density and reduced sputtering yields from high ion impact energies.

In all conditions tested, plasma exposure resulted in distinct microscopic surface morphology composed of irregular, consolidated structures. There is a lack of pits and craters normally associated with sputtering damage. Microscopic cracking is observed, which suggests the importance of thermal cycling. Plasma exposure also resulted in smoothing of the graphite surface in all cases except for the high particle fluence condition.
Overall, these changes in the surface morphology and topology indicate substantial rearrangement of material. However, definitive correlation to sublimation, melting, and redeposition processes is limited with the available data.
% implications and applicability to other platforms

The results presented here indicate some alignment of plasma-electrode interactions on ZaP-HD with processes specific to electrodes in arc discharges. Namely, there is evidence for a material recycling process that reduces net erosion and promotes electrode self-healing. This phenomenon has been observed for thermionic cathodes used in welding and plasma arc cutting tools, where evaporation and redeposition of the cathode material play a similar role. \cite{Nemchinsky_2014} This recycling process also represents an additional source of electrons that drive the large currents on ZaP-HD. These electrons are liberated through the ionization of the large flux of sublimated neutrals, which are subsequently redeposited to the cathode. This overcomes the space-charge limitation of thermionic emission. Nonetheless, it is important to note that this study focuses on the ZaP-HD cathode, and it is unlikely that such a recycling process occurs at the anode. Significant heat flux may  cause sublimation at the anode, but ionized neutrals are accelerated away from the anode due to the higher anode potential. However, future high-powered SFS Z-pinch devices will likely operate with a flowing liquid metal anode, which addresses potential issues with net erosion. \cite{Thompson_2023_FST}

As has been identified in other work, \cite{Thompson_2023_FST, Thompson_2023_POP} solutions for electrode erosion management in the SFS Z pinch benefit from the relatively small volume and mass in direct contact with plasma, and the simple geometry for component replacement. The relatively low net erosion rates reported in this study are encouraging, especially for the high plasma temperature and current densities found on ZaP-HD. This work suggests that the wealth of knowledge in operating high-powered plasma arc discharges could also be relevant for the SFS Z pinch. This provides confidence that practical solutions to manage electrode erosion in fusion devices are possible.

\begin{acknowledgments}
The authors would like to thank Ahad Ather and Helen Locke for valuable contributions to ex-situ data collection, analysis, and interpretation of surface measurements. The information, data, or work presented herein was funded in part by the National Nuclear Security Administration under Grant No. DE-NA0001860. Part of this work was conducted at the Molecular Analysis Facility, a National Nanotechnology Coordinated Infrastructure (NNCI) site at the University of Washington, which is supported in part by funds from the National Science Foundation (awards NNCI-2025489, NNCI-1542101), the Molecular Engineering $\&$ Sciences Institute, and the Clean Energy Institute. Part of this work was conducted at the Washington Clean Energy Testbeds, a facility operated by the University of Washington Clean Energy Institute. 

\end{acknowledgments}

\section*{Author Declarations}
\subsection*{Conflict of Interest}

The authors have no conflicts to disclose.

\subsection*{Author Contributions}
\noindent \textbf{Amierul Aqil Khairi:} Conceptualization (equal); Formal analysis (lead); Investigation (lead); Methodology (equal); Writing - original draft (lead); Writing - review \& editing (equal). \textbf{Uri Shumlak:} Conceptualization (equal); Funding acquisition (lead); Methodology (equal); Supervision (lead); Writing - review \& editing (equal).

\section*{Data Availability Statement}

The data that support the findings of this study are available from the corresponding author upon reasonable request.

\nocite{*}
\bibliography{aip_pop}% Produces the bibliography via BibTeX.

\end{document}